\documentclass{article}
\usepackage[T1]{fontenc} % add special characters (e.g., umlaute)
\usepackage[utf8]{inputenc} % set utf-8 as default input encoding
\usepackage{ismir,amsmath,cite,url}
\usepackage{graphicx}

\usepackage{color}

\usepackage[firstpage]{draftwatermark}
\definecolor{lightgray}{rgb}{0.9,0.9,0.9}
\definecolor{darkgray}{rgb}{0.4,0.4,0.4}
\SetWatermarkFontSize{12pt}
\SetWatermarkScale{1.1}
\SetWatermarkAngle{90}
\SetWatermarkHorCenter{202mm}
\SetWatermarkVerCenter{170mm}
\SetWatermarkColor{darkgray}
\SetWatermarkText{Late-Breaking / Demo Session Extended Abstract, ISMIR 2024 Conference}

\def\lbd{}	% Flag to use correct LBD settings in the paper, please do not modify this line

\title{Enhanced Automatic Drum Transcription via Drum Stem Source Separation}

\twoauthors
 {Xavier Riley} {Centre for Digital Music \\ {\tt j.x.riley@qmul.ac.uk}}
 {Simon Dixon} {Centre for Digital Music \\ Queen Mary University of London}

\def\authorname{X. Riley and S. Dixon}

\begin{document}

\maketitle
\begin{abstract}
Automatic Drum Transcription (ADT) remains a challenging task in MIR but recent advances allow accurate transcription of drum kits with up 5 classes - kick, snare, hi-hats, toms and cymbals - via the ADTOF package. In addition, several drum kit \emph{stem} separation models in the open source community support separation for more than 6 stem classes, including distinct crash and ride cymbals. In this work we explore the benefits of combining these tools to improve the realism of drum transcriptions. We describe a simple post-processing step which expands the transcription output from five to seven classes and furthermore, we are able to estimate MIDI velocity values based on the separated stems. Our solution achieves strong performance when assessed against a baseline of 8-class drum transcription and produces realistic MIDI transcriptions suitable for MIR or music production tasks.
\end{abstract}
\section{Introduction}\label{sec:introduction}

Automatic Drum Transcription (ADT), a sub-task of Automatic Music Transcription (AMT), offers huge potential for extracting useful data from music signals. The history of ADT methods is well summarised in the literature \cite{fitzgerald_unpitched_2006,wu_adt_review_2018}. Both cited surveys show that the history of ADT mirrors that of AMT in terms of which techniques were adopted; for example the progression from signal processing techniques to NMF based approaches to the more recent deep learning methods. However, while AMT accuracy for pitched instruments has increased steadily over time \cite{oaf,kong,yan_scoring_2024}, the accuracy for ADT methods remains far below that of, say, piano. As identified by Wu et al.\ \cite{wu_adt_review_2018}, they are affected by data issues such as small size, lack of complexity or lack of diversity (homogeneity).

Callender et al.\ \cite{callender_egmd} showed that an Onsets and Frames \cite{oaf} style ADT model which included velocity data demonstrated strong performance in user preference studies, due to the inclusion of velocity information. However, the released model appears to struggle with generalisation as MP3 encoding can negatively affect real-world transcription results\footnote{\url{github.com/magenta/magenta/issues/1876}}.

Vogl et al.\ \cite{vogl_drum_2017} introduced the use of a Convolutional Recurrent Neural Network (CRNN) which transcribed 3 drum classes (BD, SN, HH) with SOTA results on the ENST dataset. This was later expanded to 8 drum classes \cite{vogl_towards_2018}, albeit with lower accuracy. The current state of the art for ADT is ADTOF \cite{zehren_2023_high}, which uses a similar CRNN architecture but with scaled up training data. However, the output of the ADTOF model is limited to 5 classes due to the source material used as training data.

Source separation has seen an explosion of interest in recent years with the advent of deep learning models for this task. While separating drum kits from a mixture is well studied \cite{demucs}, a relatively new task is performing audio separation of individual components from a drum kit performance - for example decomposing a drum kit into kick, snare, toms, hi-hat and cymbal stems. LarsNet \cite{mezza2024toward} is one such solution, however the training data is derived entirely from drum samples produced by Logic Pro X. This lack of diversity could harm separation quality when extending to real-world examples.

Our work explores a recent open source contribution, which we refer to as "Jarredou"\footnote{\url{github.com/jarredou/models/releases/tag/aufr33-jarredou_MDX23C_DrumSep_model_v0.1}}, that has not yet been described in the scientific literature. In correspondence with the authors, we understand that this is trained on a private dataset of MIDI and rendered audio from drum-sample libraries with 21.8 hours of audio in the training set and 0.27 hours of audio reserved as a validation set. There is some repetition in the MIDI annotations as they are re-rendered using a variety of sample libraries from different providers. A full breakdown of its separation performance is planned for future work, however empirically we find that it performs well on a variety of recordings.

\section{Method}

Our method operates on solo drum kit audio (solo drums). Where the desired source is already part of a mix, we first isolate the drum part using Demucs v4\cite{demucs}. The input audio is initially normalized to a constant level using the ReplayGain algorithm\footnote{implemented in Essentia\cite{essentia}} to ensure a relatively consistent dynamic level for later processing.
We proceed to transcribe the solo drums via ADTOF to extract note locations for 5 drum classes (kick, snare, hi-hat, toms and cymbals). We also separate the solo drums using the Jarredou model to extract 6 stems, with the cymbals class is expanded to crash and ride stems.

To recover velocity information we marry the transcription data to the drum stem data as follows: first a loudness curve is computed for each stem. We use an equal loudness filter from Essentia to preserve perceptual balance, then calculate the RMS with a 1024 sample window (at a 44.1kHz sample rate) and a 10ms hop size. The RMS values are then converted to a decibel (log) scale. All six stems are then normalized to the peak dB across the group -- this allows the performance velocity to be self-consistent within a single performance, but if a reference level is known it can be substituted instead.

For each note in the transcribed MIDI, we take a 50ms window around the predicted onset time applied to the loudness curve for the corresponding stem. We then extract a MIDI velocity estimate from 0-127 by taking the maximum value of the window, scaled to the normalized dynamic range across all stems in a performance.

\subsection{Recovering additional instrument classes}

We further enhance the transcription by increasing the number of predicted classes from 5 to 7 as follows. For the cymbal onset predictions, we compare the loudness curves for the crash and ride stems and choose the maximum. However, the nature of crash cymbals means they often have a long, slow decay in amplitude which can cause incorrect classifications. To address this we introduce a heuristic by identifying significant crash cymbal peaks over the entire performance. For each crash peak, a refraction period is added which lasts until 1 second before the next peak. During this period a crash cymbal cannot be re-triggered and all cymbal hits during the period are assigned to the ride instead.

For the hi-hat stem, we observe that the loudness curve for open hi-hats decays more slowly. For each hi-hat note we take a window of the loudness curve up to the next hi-hat onset or 150ms, whichever is smaller. If the minimum loudness over the window is greater than 75\% of the maximum we assign the note to the open hi-hat class, otherwise it is assigned to the closed hi-hat class.

% yielding different results. Through experimentation, we have determined that the ADTOF is particularly sensitive to the level of the input - for example compressive normalization increases transcription accuracy for kick, snare and toms overall but tends towards higher recall and lower precision.

% Transient processing is achieved by extracting percussive components using Harmonic-Percussive Source Separation (HPSS \cite{fitzgerald_hpss}). The harmonic component is discarded and the percussive is added back into the mixture to boost the energy in transients. The result is then normalized to -12dB LUFS to increase the energy of transients even further. This processing step was found to increase ADTOF accuracy for hihats, albeit marginally.

% Finally the unaltered audio is fed to ADTOF as this was found to be most accurate for cymbals and toms.

% We combine these three transcriptions to produce a hybrid transcription, which is then paired with the source separated stems from the Jarredou model.

\begin{figure}
 \centerline{\framebox{
 \includegraphics[alt={Distribution of Estimated MIDI Velocities for MDB dataset},width=0.9\columnwidth]{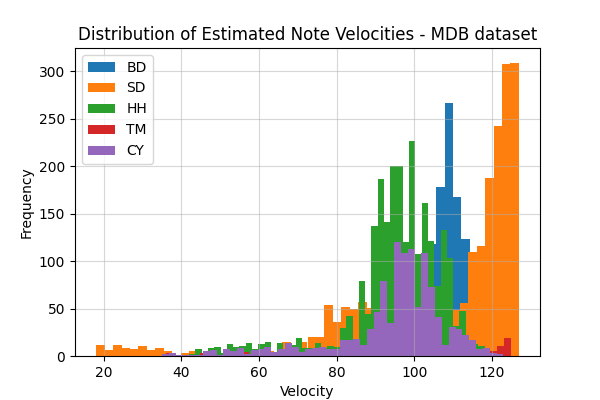}}}

 \caption{Estimated Velocity Distributions for the MDB dataset}
 \label{fig:mdb_vel}
\end{figure}

\section{Results}

The most commonly used datasets for ADT evaluation to date are ENST\cite{gillet_enst_2006}, MDB\cite{southall2017mdb} and RBMA\cite{vogl_drum_2017} which are all publicly available. Table \ref{tab:5-class} shows results for transcribing 5 drum classes (BD, SN, HH, CY, TOMS) following the framework of Zehren et al.\  \cite{zehren_2023_high}. The underlying transcription accuracy is essentially the same as that of ADTOF, with the only difference being our treatment of zero velocity notes which are omitted from the evaluation.

We also include results for the 8-class transcription task in Table \ref{tab:8-class}. Our model does not attempt to predict the relatively rare Cowbell class (MIDI 56) so a direct comparison is not possible, however we include baseline results \cite{vogl_towards_2018} for reference. These show strong increases in performance over the baseline for the MDB and ENST datasets (12\% and 10\%), however RBMA is 2\% lower. We believe this is due to the electronic drum sounds used in this dataset which are outside of the domain of acoustic drums.

\begin{table}
 \begin{center}
 \begin{tabular}{|l|l|l|l|l|}
  \hline
  Method & Dataset & Recall & Precision & F-measure \\
  \hline
  Ours & MDB & 0.89 & 0.89 & 0.89 \textit{(0.87)} \\
  Ours & RBMA$\dag$ & 0.62 & 0.66 & 0.63 \textit{(0.65)}\\
  Ours & ENST & 0.81 & 0.91 & 0.85 \textit{(0.84)} \\
  \hline
 \end{tabular}
\end{center}
 \caption{Results for 5-class transcription accuracy across 3 datasets. Onset-only with 50ms tolerance via \texttt{mir\_eval}. Obelisk (\dag) indicates drums were isolated using \cite{demucs}. Original ADTOF results are shown in parentheses}
 \label{tab:5-class}
\end{table}

\begin{table}
 \begin{center}
 \begin{tabular}{|l|l|l|l|l|}
  \hline
  Method & Dataset & Recall & Precision & F-measure \\
  % \hline
  % \cite{vogl_towards_2018} & MDB  & - & - & 0.72 \\
  % \cite{vogl_towards_2018} & RBMA & - & - & 0.58 \\
  % \cite{vogl_towards_2018} & ENST & - & - & 0.65 \\

  \hline
  Ours & MDB & 0.84 & 0.84 & 0.84 \textit{(0.72)} \\
  Ours & RBMA$\dag$ & 0.55 & 0.60 & 0.56 \textit{(0.58)} \\
  Ours & ENST & 0.72 & 0.81 & 0.76 \textit{(0.65)} \\
  \hline
 \end{tabular}
\end{center}
 \caption{Results for 8-class transcription accuracy. See Table \ref{tab:5-class} for details. Results in parentheses show baseline results for 8-class transcription from Vogl et al.\ \cite{vogl_towards_2018}.}
 \label{tab:8-class}
\end{table}

The predicted velocity distributions for the MDB dataset are shown in Figure \ref{fig:mdb_vel}. The other datasets are omitted but show similar distributions. This illustrates that our method produces a range of normally distributed velocities for each class. The concentration of snare at high volumes is likely a result of the equal loudness curves favouring mid-range frequencies. Our method allows for these components to be scaled individually to improve balance, if necessary.

\section{Conclusions}

In this work we demonstrate a method of combining an ADT model (ADTOF) with a drum stem source separation model. This combination allows us to estimate velocities and perform additional levels of classification while retaining a high degree of transcription accuracy. We intend to use this in future for dataset production workflows to enhance ADT further.

\section{Acknowledgments}
XR is a research student at the UKRI Centre for Doctoral Training in Artificial Intelligence and Music, supported by UK Research and Innovation [grant number EP/S022694/1].

% For bibtex users:
\bibliography{ISMIR2024_lbd}

\end{document}